\documentstyle[prl,aps,multicol,epsf]{revtex}

\newcommand{\lineup}{\raisebox{5mm}[0mm][0mm]{\makebox[0mm][l]%
{\epsfxsize=9.0cm\epsfbox{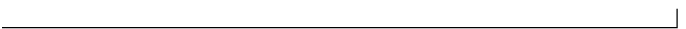}}}}
\newcommand{\linedown}{\raisebox{3mm}[0mm][0mm]{\makebox[0mm][l]%
{\hspace{9.0cm}\epsfxsize=9.0cm\epsfbox{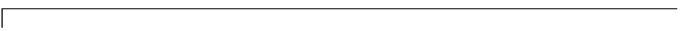}}}}
\newcommand{\half}{\frac{1}{2}}
\newcommand{\E}{\varepsilon}

\newcommand{\lts}{
   \parbox{1mm}{
      \setlength{\unitlength}{1mm}
      \begin{picture}(1,12)
         \thinlines
         \put(0.5,-0.2){\line(0,1){11}}
      \end{picture}
   }
}

\newcommand{\be}{\begin{equation}}
\newcommand{\ee}{\end{equation}}
\newcommand{\bea}{\begin{eqnarray}}
\newcommand{\eea}{\end{eqnarray}}

\newcommand{\weiter}{\nonumber \\ & & }
\newcommand{\nicht}[1]{}

\newlength{\pcm}
\newlength{\pmm}
\newcommand {\GB} {\,\epsfxsize=1.2\pcm \parbox{1.2\pcm}{\epsfbox{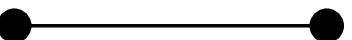}}\,}

\newcommand {\GH} {\,\epsfxsize=0.8\pcm \parbox{0.8\pcm}{\epsfbox{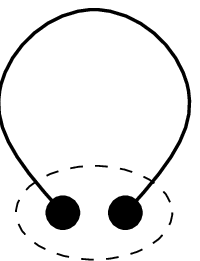}}\,}
\newcommand {\GI} {\,\epsfxsize=2\pcm \parbox{2\pcm}{\epsfbox{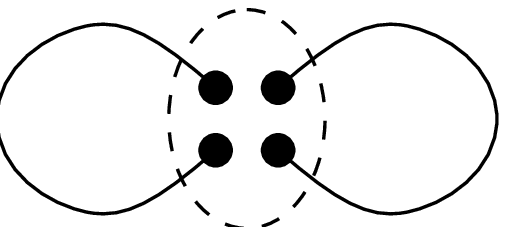}}\,}
\newcommand {\GJ} {\,\epsfxsize=1.3\pcm \parbox{1.3\pcm}{\epsfbox{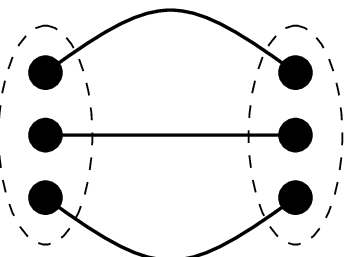}}\,}

\newcommand {\GM} {\,\epsfxsize=1.5\pcm      
\parbox{1.5\pcm}{\epsfbox{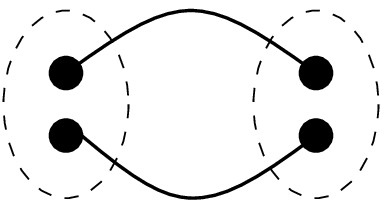}}\,}

\newcommand {\GO} {\,\epsfxsize=0.4\pcm \parbox{0.4\pcm}{\epsfbox{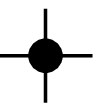}}\,}
\newcommand {\GP} {\,\epsfxsize=2.5\pcm \parbox{2.5\pcm}{\epsfbox{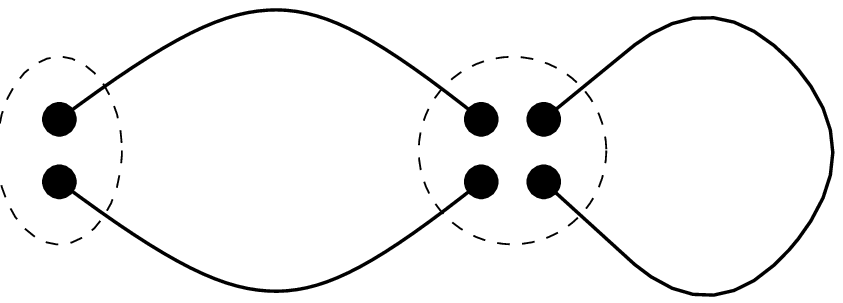}}\,}
\newcommand {\GPpsub} {\,\epsfxsize=2.5\pcm \parbox{2.5\pcm}{\epsfbox{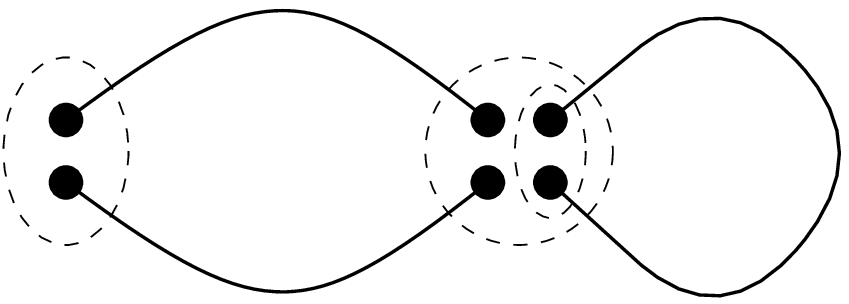}}\,}

\newcommand {\GW} {\,\epsfxsize=1.2\pcm
\parbox{1.2\pcm}{\epsfbox{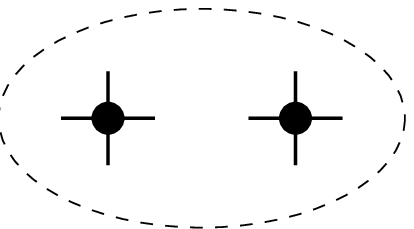}}\,}
\newcommand {\GX} {\,\epsfxsize=2.4\pcm
\parbox{2.4\pcm}{\epsfbox{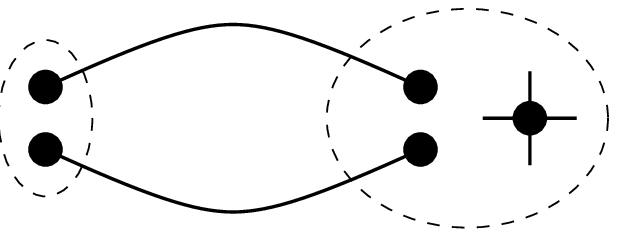}}\,}

\newcommand {\GZ} {\,\epsfxsize=1.1\pcm
\parbox{1.1\pcm}{\epsfbox{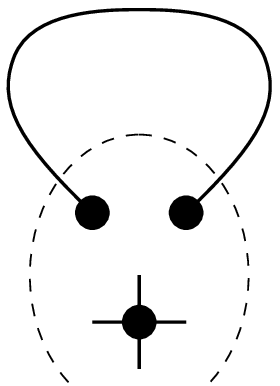}}\,}

\title{Scaling of Selfavoiding Tethered Membranes:\\
2-Loop Renormalization Group Results}
\author{\centerline{Fran\c cois David$^{\dagger}$ and Kay J\"org Wiese}
\newline
\centerline{Service de Physique Th\'eorique, C.E.A. - Saclay, 
91191 Gif-sur-Yvette Cedex, France}
}
\date{\today}

\begin{document}
\maketitle
\begin{abstract}

The scaling properties of selfavoiding polymerized membranes are 
studied using renormalization group methods.
The scaling exponent $\nu$ is calculated for the first time at two loop
order.
$\nu$ is found to agree with the Gaussian variational estimate for large
space dimension $d$ and to be close to the Flory estimate for $d=3$.
\end{abstract}

\begin{multicols}{2}
\noindent
The statistical properties of polymerized flexible membranes are interesting
and still poorly understood \cite{r:Jerus}.
These objects arize 
either in a collapsed (fractal dimension $d_f=3$),
a crumpled swollen  ($2<d_f<3$) or a flat ($d_f=2$) phase.
The physical properties of such membranes in three dimensions 
can be studied by experiments and computer-simulations.
Most of the simulations find a flat phase \cite{r:Abr&al89,r:AbrNel90,r:GreMur90,r:GrestPetsche94,Gompper}, 
due to an induced effective curvature term (stiffness)
of the membrane \cite{r:AbrNel90}.
A swollen phase with fractal dimension 
near to the Flory-prediction $d_f=2.5$ has been found 
by exactly balancing curvature terms with a long-range 
repulsive interaction, thus interpreting the swollen phase
as a tricritical point \cite{r:GrestPetsche94}.
The experimental results are contradictory. In \cite{r:graphitefrac} but not 
in \cite{r:graphiteflat} a swollen phase is found.   

An analytical approach inspired from polymer theory \cite{r:CloJan90}, 
which relies on renormalization group and $\E$-expansion methods,
was initiated in \cite{r:AroLub87,r:KarNel87}, where it was
used to perform calculations at 1-loop order.
Its consistency to all orders in perturbation theory has been established in \cite{r:DDG3}.
In this letter we present the first application of this method to 
2-loop calculations and discuss the results obtained by this approach.

The membrane is modeled by a continuum model \`a la Edwards:
the embedding of the $D$-dimensional membrane in $d$-dimensional bulk space
is described by the mapping $x\in R^D\to \vec r(x)\in R^d$.
The renormalized Hamiltonian is
\begin{equation} \label{e:Ham}
\!\!\!\!{\cal H}[\vec r]= \frac{Z}{2}\int_x\!\big(\nabla \vec r(x)\big)^2
+ b Z_b \mu^\varepsilon \int_x\!\int_y\!
\delta^d\big (\vec r(x)-\vec r(y)\big ) \ ,
\end{equation}
where $b$ is the dimensionless renormalized coupling constant, $\mu$ the
renormalization momentum scale and $\E=2D-d(2-D)/2$.
Physical quantities are calculated perturbatively in $b$.
Direct calculations for $D=2$ for membranes are not possible, since perturbation
theory is singular; $D$ and $d$ have therefore to be treated as continuous variables
and an $\varepsilon$-expansion must be performed. 
The renormalization factors
$Z(b,\varepsilon)$ and $Z_b(b,\varepsilon)$ are introduced in order to
subtract the short-distance divergences which appear as poles in $\varepsilon$
at the critical dimension $\varepsilon=0$.

At order $b^n$ one has to to calculate the expectation
value with respect to the free theory ($b=0$) of $n$ bilocal operators,
henceforth called dipoles:
$_x\GB_y =\delta^d(\vec r(x)-\vec r(y))$
integrated over the whole membrane.
Short-distance divergences occur when dipole end-points approach 
each other.
The most important tool to deal with these divergences 
is the multilocal operator product expansion (MOPE)
\cite{r:DDG3},
which describes all possible contractions of dipoles to the operators
marginal at $\E=0$.
Power counting shows that there are only two such operators:
the dipole operator and the local operator:
$\half (\nabla r(x))^2=\GO_x$.
For instance, the contraction of a dipole to a point generates $\GO$
with the MOPE coefficient
\begin{equation}
\label{e:mopecoef}
\bigg( \vphantom{\hbox{\GH}}_x\GH_y \bigg| \GO\bigg)=
-\frac1{2D} |x-y|^{\varepsilon-D} \ .
\end{equation}
The integral over the relative distance of the two points is UV-divergent.
As in \cite{r:WieseDavid95} we use the minimal subtraction scheme to subtract these divergences. Introducing an IR-cutoff 
 $L\propto\mu^{-1}$ the Feynman-diagram becomes
\begin{equation}
\label{e:MOPEL}
\bigg< \GH \bigg| \GO\bigg>_L = \int_{\mbox{\scriptsize
all distances}\atop\mbox{\scriptsize smaller than $L$}} \bigg( \GH \bigg| \GO\bigg) \ .
\end{equation}
Our strategy is to keep $D$ fixed and to 
expand (\ref{e:MOPEL}) as a Laurent series in
$\varepsilon$, which here starts at $\varepsilon^{-1}$.
Denoting by $\big<\ \big|\ \big>_{\varepsilon^p}$ the term of order
$\varepsilon^p$ of $\big<\ \big|\ \big>_{L=1}$, the counterterms are
chosen to have pure poles in $\varepsilon$, and for instance the first
counterterm corresponding to (\ref{e:mopecoef}) is
$ \displaystyle \big<\GH\big|\GO\big>_{\varepsilon^{-1}}= -\frac1{2D} \frac1\E$.
\end{multicols}
\noindent%
\lineup
Extending this analysis along the lines of \cite{r:DDG3}\ leads to the
following results (details will be given elsewhere \cite{r:DaWi96b}).
To second order, the counterterms which render the theory finite are found
to be
\begin{eqnarray} \label{e:Z}
Z&=& 1-\bigg< \GH \bigg| \GO \bigg>_{\E^{-1}} {b}
+ \Bigg[  \bigg< \GI \bigg| \GO \bigg>_{\E^{-2},\E^{-1}}
-2\bigg< \GH \bigg| \GO \bigg>_{\E^{-1}}
\bigg< \GZ \bigg| \GO \bigg>_{\E^{-1},\E^0}
\weiter
\qquad \qquad \qquad \qquad \qquad \ + \bigg< \GH \bigg| \GO \bigg>_{\E^{-1}}^2
\bigg< \GW \bigg| \GO
\bigg>_{\E^0} 
 - 2 \bigg< \GM \bigg| \GB \bigg>_{\E^{-1}} \bigg< \GH \bigg| \GO \bigg>_{\E^0}  \bigg] \frac{b^2}{2!}
\end{eqnarray}

\begin{eqnarray} 
\label{e:Zb}
Z_b&=&1+2\bigg< \GM \bigg| \GB \bigg>_{\E^{-1}}\, \frac b {2!} 
- \bigg[
 4\bigg< \GJ \bigg| \GB \bigg>_{\E^{-2},\E^{-1}} 
-12\bigg< \GM \bigg| \GB \bigg>_{\E^{-1}} \bigg< \GM \bigg| \GB \bigg>_{\E^{-1},\E^0}
\weiter
+12 \bigg< \GP \bigg| \GB \bigg>_{\E^{-2} , \E^{-1}} 
-12 \bigg< \GH \bigg| \GO \bigg>_{\E^{-1}} \bigg< \GX \bigg| \GB \bigg>_{\E^{-1},\E^0}  \bigg] \frac{b^2}{3!}
\end{eqnarray}
%
\begin{multicols}{2}\narrowtext
\noindent
\linedown
Following \cite{r:DDG3},
the renormalization group $\beta$-function and the anomalous scaling
dimension $\nu$ of $\vec r$ are obtained from the  variation of the 
coupling constant 
and the field with respect to the renormalization scale $\mu$,
keeping the bare couplings fixed. They are written in terms of
$Z$ and $Z_b$ as 
\begin{eqnarray}
\label{e:beta}
\beta(b) &=&
\frac{-\E b} {1+ b\frac{\partial}{\partial b } \ln Z_b +
\frac{d}2 b \frac{\partial}{\partial b} \ln Z}
\\
\label{e:nu}
\nu (b) &=&
\frac{2-D}2 -\half \beta(b) \frac{\partial}{\partial b} \ln Z 
\end{eqnarray}
%
In order to compute these coefficients, we apply our methods of
\cite{r:WieseDavid95}, which must be extended in order to deal with the
double poles in $\E^{-2}$ from subdivergences. 
We demonstrate the method using the example of the counterterm associated with the
diagram ${\cal G}=\GP$
\begin{equation}
\bigg< \GP \bigg| \GB \bigg>_L
=L^{2\E}\Big[{c_2\over\E^2}+{c_1\over\E}+{\cal O}(\E^0)
\Big]\label{e:huhu}
\end{equation}
A subdivergence occurs when the single dipole to the right of the diagram
${\cal G}$ is contracted to a point.
According to the MOPE of \cite{r:DDG3}, when this contraction is performed
{\it first}, the MOPE coefficient factorizes as
\begin{eqnarray}
\bigg( \GPpsub \bigg| \GB \bigg) &\approx& \bigg( \GH \bigg| \GO \bigg) \times
\weiter
\quad \times	 \bigg( \GX \bigg| \GB \bigg)
\label{e:MOPECT}
\end{eqnarray}
This implies that the double pole of (\ref{e:huhu}) is the same 
as the double pole appearing in the product of the
counterterms associated with the two diagrams on the r.h.s.\ of (\ref{e:MOPECT})
\begin{eqnarray}
\label{e:secondcoupling}
&&\hspace {-.7cm} L^{2\E} \left[ {c_2\over\E^2} + {\tilde c_1 \over \E} +{\cal O}(\E^0) \right]
\nonumber \\
&& =  \half \bigg< \GH \bigg| \GO \bigg>_{L}
\bigg< \GX \bigg| \GB \bigg>_{L} 
\end{eqnarray}
The factor $1/2$ comes from the nested integration
\cite{r:DuHwKa}, arizing from the fact that 
the double pole is given by the r.h.s.\ of (\ref{e:MOPECT}), integrated
with the restriction $\cal R$ that the distance in $\GH$ is smaller than all the
distances in $\GX$.
As a consequence, we can extract the difference of the residues of the single poles, $c_1-\tilde c_1$,
by subtracting to the l.h.s.\
of (\ref{e:huhu}) a counterterm proportional to the r.h.s.\ of (\ref{e:MOPECT}),
{\em restricted to the domain ${\cal R}$}. 
In fact, this combination has to be calculated,
see (\ref{e:Zb}). We get:
\end{multicols}
\widetext
\noindent \lineup \vspace{-6mm}
\begin{eqnarray} 
\label{e:c_one}
&L^{2\E}&\Big[{c_1-\tilde c_1\over\E} +{\cal O}(\E^0)\Big]
=
\int_{{\rm distances}\atop \le L}
\!\!
\bigg( \GP \bigg| \GB \bigg) 
-\int_{\left\{{{\rm distances}\atop \le L}\right\}\cap{\cal R}}
\bigg( \GH \bigg| \GO \bigg)\times
\bigg( \GX \bigg| \GB \bigg)
\end{eqnarray}
The careful reader will have remarked that (\ref{e:secondcoupling})
and the last term in  (\ref{e:c_one}) are calculated with a slightly 
different regularization prescription. The difference however is 
of order $\E^0$ and thus does not change the result \cite{r:DaWi96b}. 
$c_1-\tilde c_1$ is now extracted by applying
${\small L}{\partial\over\partial L}$ to the r.h.s.\ of (\ref{e:c_one}) and
taking the limit $\E\to 0$.
We then obtain
\begin{eqnarray}
c_1-\tilde c_1&=&{L\over 2}\,\lim_{\E \to 0} \left[
\int_{\left\{ {\rm largest}\atop{{\rm distance}=L} \right\} }
\bigg( \GP \bigg| \GB \bigg)\right.
-\left.
\int_{\left\{{{\rm largest}\atop{\rm distance} =L}\right\}\cap{\cal R}}
\bigg( \GH \bigg| \GO \bigg) \times \bigg( \GX \bigg| \GB \bigg)
\right] 
\label{e:cone}
\end{eqnarray}
\begin{multicols}{2} \narrowtext
\noindent \linedown
This integral is locally finite and 
can be reduced to an integral over five independent distances
between pairs of points.
We evaluate it by applying and extending the numerical techniques of
\cite{r:WieseDavid95}.
The main difficulty thereby comes from the fact that although convergent,
the integral (\ref{e:cone}) has integrable singularities, which have to
be removed by suitable variable transformations and mappings between
domains of integration.
Finally the integrand has large variations in some small subdomains
and can only be integrated by a genuine adaptive Monte Carlo integration.
For analytical and numerical details we refer the reader to
\cite{r:WieseDavid95,r:DaWi96b}.

The other diagrams appearing in (\ref{e:Z}) and (\ref{e:Zb}) are calculated similarly.
Having performed all numerical calculations
($\sim 10^3$h CPU on a WorkStation), the renormalization-group 
%
%
functions $\beta(b)$ and $\nu(b)$ can be calculated.
$\beta(b)$ has a non-trivial IR-fixed point for $b=b_c>0$ (i.e.\ $\beta(b_c)=0$ 
and $\beta'(b_c)>0$).\
The full dimension of the field $\nu$ at this critical point, $\nu(b_c)$ is
a function of $D$ and $\E$:
\be \label{e:nuuu}
	\nu(D,\E) = \frac{2-D}2 + \nu_1(D) \E + \nu_2(D) \E^2 + {\cal O} (\E^3)
\ee
\begin{figure}[h] \label{f:nu12}
\centerline{
\epsfxsize=8.7cm \parbox{8.7cm}{\epsfbox{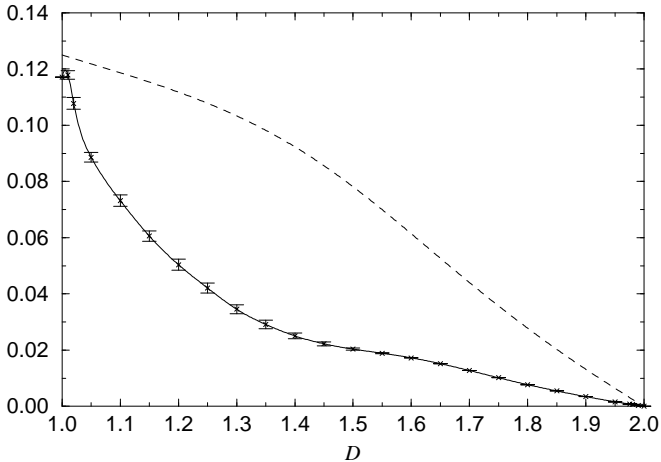}}
}
\caption{The functions $\nu_1(D)$ (dashed line) and $\nu_2(D)$ (solid line).
The latter is given with the statistical error. }
\end{figure} 
The coefficients are plotted in figure 1. 
Naively setting 
$D=2$ and $d=3$, i.e.\ $\E=4$ 
in equation (13), yields the wrong result 0. To obtain the 
correct result, one has to redevelop this expansion around any point
$(D_0, d_0=\frac{4D_0}{2-D_0})$ on the critical curve ($\E=0$).
We use a generalization of the methods introduced in \cite{Hwa}.
Note that the expansion (\ref{e:nuuu}) 
is exact in $D$ and of order 2 in $\E$, thus can be expanded up to 
order 2 in $D-D_0$ and $\E$.
Given any invertible transformation 
$\{ x,y \} = \{x(D,\E), y(D,\E) \}$, one can express $D$ and $\E$ as function
of $x$ and $y$ and re-expand up to order 2 in $x$ and $y$.
The goal is to find a set of variables, such that the estimate for $\nu$ depends
the least on the choice of 
the expansion point on the critical curve and which 
reproduces well the known results in the following cases:
$D=1$, $D=d$ and $d\to \infty$ (see below).
The sets $\{ D,D_c(d)=\frac{2d}{4+d}\}$ and $\{ D_c(d),\E \}$ have been found to
be good choices. 
The plot in figure 2 shows the value of $\nu$ as a function of the expansion
point $\{D,d_c(D)=4D/(2-D)\}$ on the critical curve. The prediction at 1-loop order (dashed line) 
is essentially independent of the expansion point. At 2-loop order the 
estimate starts from the 1-loop result at $D=2$, grows until it
reaches a plateau around $D=1.5$ and then grows rapidly again.
This is a general feature of this kind of expansion and can well be
studied by applying the same method to the Flory-estimate
$\nu_{\mbox{\scriptsize Flory}}=
(2+D)/(2+d)$. 
In this case 
the plateau becomes flatter if one goes to higher orders but 
does not cover the whole range, i.e.\ the expansion is not
convergent for all $D$. 
To extract $\nu$ from figure 2, one uses the  maximum and
the minimum of the 
\begin{figure}[h] \label{f:extrapole}
\centerline{
\epsfxsize=8.7cm \parbox{8.7cm}{\epsfbox{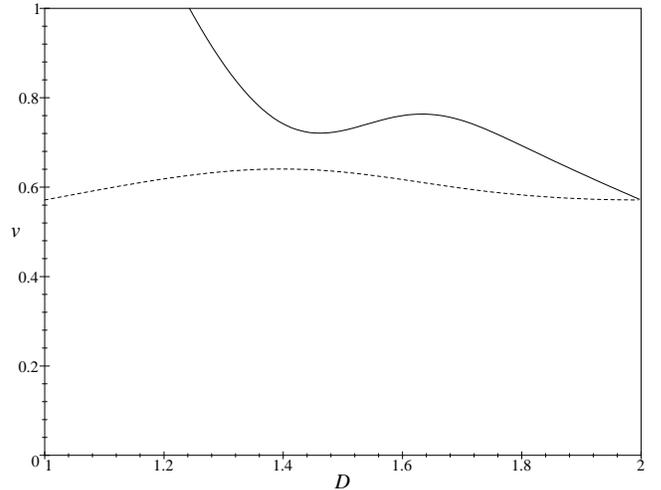}}
}
\caption{Extrapolation of $\nu$ via (\protect\ref{e:nuuu}) in $D_c(d)$ and $\E$ for membranes $D=2$ in $d=3$. The first order results are
given by the dashed line, the second order by the solid line.}
\end{figure} 
\noindent plateau.  Their mean is an estimate for 
$\nu$, their difference an estimate of the error in {\em this} expansion scheme.
We get $\nu = 0.74\pm 0.02$. It must be emphasized that different sets 
of variables yield different values for $\nu$. 

One possibility  to further improve the extrapolation is to 
develop $\nu d$ or $\nu (d+2)$.
The first is interesting, as it calculates corrections around the estimate
predicted by a Gaussian variational ansatz\cite{Guitter variational},
$\nu_{\mbox{\scriptsize var}} =2D/d$:
\be 
\nu(b_c) d = 2D + \left( \beta(b) \frac{\partial}{\partial b} \ln Z_b(b) 
		 \right)\lts_{b=b_c}
\ee
The smaller $\ln(Z_b)$ is, the more accurate the
expansion becomes. 
At 1- and 2-loop order $\ln (Z_b)$ vanishes like $\exp (-\mbox{const}/d)$
for large $d$.
We argue that this persists to any order in perturbation theory.
For large $d$, $\nu_{\mbox{\scriptsize var}}$ thus becomes exact.

\begin{figure}[h] \label{f:extrapole2}
\centerline{
\epsfxsize=8.7cm \parbox{8.7cm}{\epsfbox{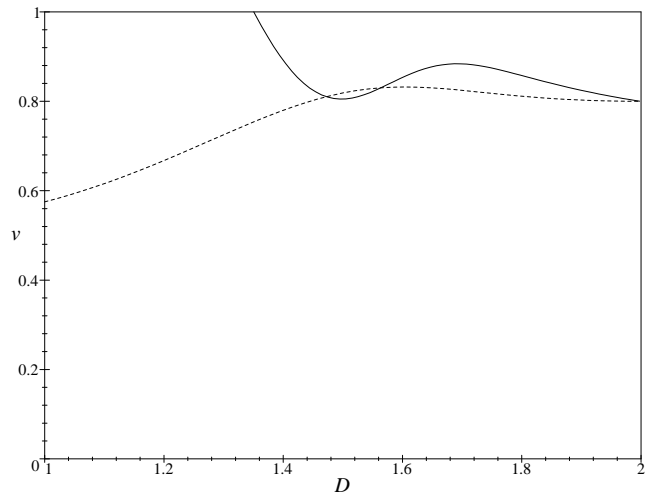}}
}
\caption{Extrapolation of $\nu$ via (\protect\ref{nudp2}) in $d$ and $\E$ for membranes $D=2$ in $d=3$. The first order results are
given by the dashed line, the second order by the solid line.}
\end{figure}  
The second possibility is a systematic 
expansion around the Flory-estimate
$\nu_{\mbox{\scriptsize Flory}} =(D+2)/(d+2)$:
\be \label{nudp2}
\nu(b_c) (d+2) = D+2 + \left( \beta(b) \frac{\partial}{\partial b} \ln\!\left( \frac{Z_b}{Z} \right) \right)\lts_{b=b_c} 
\ee
The Flory-estimate is excellent
for polymers and one hopes \cite{Hwa} that it is 
also good for membranes. Therefore the
corrections should be small.
Such an extrapolation is given in figure 3. 
The estimate for $\nu$ is slightly larger than that in figure 2. 

The results of a 2-loop extrapolation for $\nu$ are presented in figure 4
for membranes ($D=2$) in $d$-dimensions ($2\le d\le 20)$.
\begin{figure}[h] 
\centerline{
\epsfxsize=8.7cm \parbox{8.7cm}
{\epsfbox{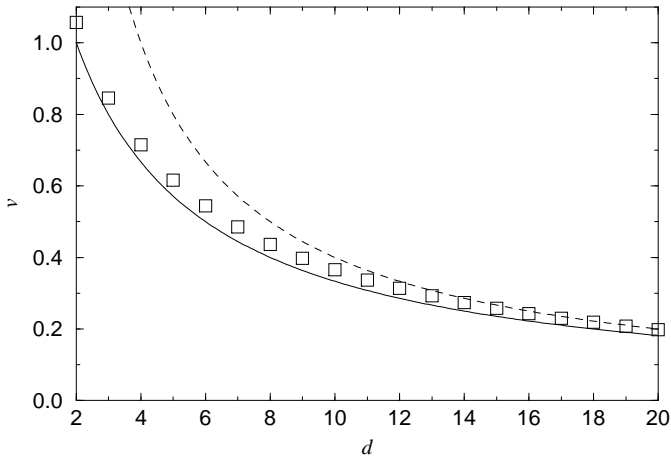}}}
\caption{Extrapolation of the 2-loop results in $d$ and $\E$ for membranes
$D=2$ in $d$ dimensions, using (\protect\ref{nudp2}) (squares). The solid line is the prediction made
by Flory's theory, the dashed line by the variational ansatz.}
\end{figure} 
We see that for $d\to\infty$ the prediction of the Gaussian variational
method becomes exact, as argued above.
For small $d$, the prediction made by Flory's argument is close to our
results.
This is a non-trivial result, since the membrane case corresponds to
$\E=4$ and in comparison with polymers in $d=3$, where $\E=1/2$,  the 2-loop corrections
were expected to be large.
In fact we have found that they are small when one expands around
the critical curve $\E=0$ for an adequate range of $D\sim 1.5$ (depending
slightly on $d$ and on the choice of variables) and a suitable
choice of extrapolation variables.
In this case the 2-loop corrections are even smaller that the 1-loop
corrections and allow for more reliable extrapolations to $\E=4$.

In conclusion, we have presented here the first renormalization
group calculation at 2-loop order
for self-avoiding flexible tethered membranes.
In order to improve these results, one should understand  if the
plateau phenomenon observed at 2-loop order persists to higher orders, and
one should
control the general large order behavior of perturbation theory for this
model. 
Another important issue is whether the IR-fixed point studied here is stable
towards perturbation by bending rigidity.
Indeed, it has been argued \cite{r:AbrNel90} that for small enough $d$ this might destabilize
the crumpled phase and explain why numerical simulations in $d=3$ normally see
a flat phase.

\acknowledgments 
We thank E. Guitter and J. Zinn-Justin for useful discussions and 
E. Guitter and Chitra for a careful reading of the manuscript.

\end{multicols}
\end{document}